\documentclass[12pt]{article}
\usepackage[ngerman,english]{babel}
\usepackage{amsmath,amssymb,amsfonts,textcomp}
\usepackage{calc}
\usepackage{hyperref}
\hypersetup{colorlinks=true, linkcolor=blue, filecolor=blue, pagecolor=blue, urlcolor=blue}
\makeatletter
\newcommand\ps@Standard{%
\renewcommand\@oddhead{}%
\renewcommand\@evenhead{}%
\setlength\paperwidth{8.5in}\setlength\paperheight{11in}\setlength\voffset{-1in}\setlength\hoffset{-1in}\setlength\topmargin{1in}\setlength\headheight{12pt}\setlength\headsep{0cm}\setlength\footskip{12pt+0.461in}\setlength\textheight{11in-1in-0.5in-0cm-12pt-0.461in-12pt}\setlength\oddsidemargin{1.25in}\setlength\textwidth{8.5in-1.25in-1.25in}
\renewcommand\thepage{\arabic{page}}
\setlength{\skip\footins}{0.0398in}\renewcommand\footnoterule{\vspace*{-0.0071in}\noindent\textcolor{black}{\rule{0.25\columnwidth}{0.0071in}}\vspace*{0.0398in}}
}
\makeatother

\begin{document}

\Large
\centerline{\bf Einstein's Apple and Relativity's Gravitational Field}

\medskip

\centerline{\bf by}

\medskip

\centerline{\bf{Engelbert Levin Sch\"ucking}}

\bigskip

\section{The First Principle of Equivalence}

\subsection{Einstein's Apple}

In 1907 Johannes Stark, editor of \textit{The Yearbook of
Radioactivity and Electronics,} asked Albert Einstein to write a review
of relativity theory for its volume 4. It was an unusual request in
Germany's academic world to charge a Second Class Expert at the Swiss
Patent Office with the task of a senior professor, to survey recent
developments in his field. Einstein's article\footnote{\ Einstein,
Albert. ``The Collected Papers of Albert Einstein'', Volume 2. English
Translation. Anna Beck, Translator; Peter Havas, Consultant. Princeton
University Press. Princeton NJ 1989. p. 252.} \textit{On the Relativity
Principle and the Conclusions drawn from it} was the result. The last
nine pages of this 52 page paper bore the title \textit{Principle of
Relativity and Gravitation}. These pages did not review published
material; they laid the foundation to Einstein's greatest and most
original contribution to science, his theory of gravitation. Fifteen
years later he called the epiphany that inspired him \textit{der
gl\"ucklichste Gedanke meines Lebens}, the happiest thought of my
life\footnote{\ Einstein, Albert.Ref. 1, p. 265.}. In the history of
science it is referred to as Einstein's first principle of equivalence
and I call it Einstein's Apple. In a speech given in Kyoto, Japan, on
December 14, 1922 Einstein remembered his experience\footnote{\ Pais,
Abraham, ``Subtle is the Lord''. Oxford University Press, New York NY
1982. p. 179.}:

\lq\lq I was sitting in a chair in my patent office in Bern.
Suddenly a thought struck me: if a man falls freely, he would not feel
his weight. I was taken aback. The simple thought experiment made a
deep impression on me. It was what led me to the theory of gravity.\rq\rq

This was an unusual vision in 1907. He had not been watching
the antics of orbiting astronauts on television, sky{}-diving clubs did
not yet exist, and platform diving was not yet a sport's category of
the freshly revived Olympic Games. How could this thought have struck
him? Had he just been dealing with patent applications covering the
safety of elevators?

\subsection{A Patent or a Prize as Inspiration?}

Three years earlier, in 1904, the Otis Elevator Company
installed in Chicago, Illinois, the first gearless traction electric
elevator apparatus, that was of the direct drive type, known as the
``1:1 elevator''. This first modern electric elevator made its way to
Europe where, on Z\"urich's Bahnhof Strasse and elsewhere in
Switzerland, buildings went up that needed elevators. It would have
been natural for Director Friedrich Haller at the Swiss Patent Office
in Bern to put applications involving electro{}-mechanical machinery on
the desk of Einstein, his expert 2\textsuperscript{nd} class with
expertise in electromagnetism.

We have one patent application with Einstein's comments
written just a week after he had finished his last section on
Gravitation for Stark's Yearbook. It challenges Germany's electric
giant AEG and begins in his neat
handwriting\footnote{\foreignlanguage{ngerman}{\ }\foreignlanguage{ngerman}{Fl\"uckinger,
Max, ``Albert Einstein in Bern''. Verlag Paul Haupt Bern 1974. p. 63.}}:

\lq\lq The patent claim is incorrect, vague and obscurely
redacted.\rq\rq

For 7 years, from 1902 to 1909, Einstein reviewed an estimated
two-thousand patents. These reviews of patent applications probably
constituted the bulk of Einstein's writings in his most productive
years. Comparing them with his papers on physics and searching them for
clues to his great discoveries might give fascinating insights into the
working of his mind. Unfortunately, this attempt at finding a clue to
the happiest thought of his life is doomed to failure. The AEG patent
application is the only one extant from those years he worked at the
patent office. The Swiss bureaucracy did destroy all other examples of
Einstein's expert opinions. We shall probably never know how he got the
inspiration to his first principle of equivalence.

However, there was one suggestion for the principle of
equivalence that went to the heart of the theory. It concerned the
apparent enigmatic equality of inertial and passive gravitational mass.
In 1906 the Academy of Sciences in Goettingen had offered the Beneke
Preis for proving this equality by experiment and theory through an
advertisement in the \textit{Physikalische Zeitschrift}. Since this
Journal reached practically all German speaking physicists Einstein may
have seen the offer. Two months after his Jahrbuch article Einstein
published his \textit{Ma\-schin\-chen} in this Journal.

The Baron Roland E\"otv\"os, the only entry, won
three{}-fourths of this prize (3,400 of 4,500 Marks); only
three{}-fourths, because he had just done experiments and had not
attempted a theoretical
explanation\footnote{\foreignlanguage{ngerman}{\ }\foreignlanguage{ngerman}{Runge,
Carl, ``G\"ottinger Nachrichten No.1, p.37{}-41
(1909).}\foreignlanguage{ngerman}{ }}. It has been claimed that
Einstein alone was at that time aware of the importance of the equality
of masses. Einstein's biographer Leopold Infeld wrote:

\lq\lq No one in our century, with the exception of Einstein, wondered about
this law any longer.\rq\rq \footnote{\ Infeld, Leopold, ``Albert Einstein'',
Charles Scribner's Sons , New York 1950, p. 47.}

Reading Runge's Prize Award one gets the clear idea that
the equality of the two masses was at that time the foremost question
for theoreticians in G\"ottingen like Hilbert, Minkowski, Klein, Voigt,
Schwarzschild, Runge, Wiechert, and Abraham, when Kaufmann carried out
his experiments on the mass of high energy electrons. Comparing the
Beneke Prize for 1906 with today's prize money, one may be justified to
call the prize for the equality of the masses the $\$$64,000 question.
If Einstein thought he had the answer to this question, why did he not
compete? I shall come back to this question.

\subsection{The First Principle of Equivalence}

In his 1907 review paper Einstein formulated his principle of
equivalence for the first time. He
wrote\footnote{\foreignlanguage{ngerman}{\ }\foreignlanguage{ngerman}{Einstein,
Albert. Ref.1, p. 302.}}:

\lq\lq We consider two systems $ \Sigma_1 $ and $ \Sigma_2 $
in motion. Let $ \Sigma_1 $ be
accelerated in the direction of its $X$-axis, and let $\gamma$ be the
(temporally constant) magnitude of that acceleration.
$ \Sigma_2 $ shall be at rest, but it shall be located in a
homogeneous gravitational field that imparts to all objects an
acceleration $ - \gamma $ in the direction of the $X$-axis.\rq\rq

The next sentence contains the principle of equivalence:

\lq\lq As far as we know, the physical laws with respect to
$ \Sigma_1 $ do not differ from those with respect to
$ \Sigma_2 $; this is based on the fact that all bodies are
equally accelerated in the gravitational field.\rq\rq

[ It was this last fact that had \ prompted Sir Hermann Bondi to the
observation \lq\lq If a bird-watching physicist falls off a cliff, he
doesn't worry about his binoculars, they fall with him.\rq\rq]

Einstein continued:

\lq\lq At our present state of experience we have thus no reason to
assume that the systems $ \Sigma_1 $ and
$ \Sigma_2 $ differ from each other in any respect, and in
the discussion that follows, we shall therefore assume the complete
physical equivalence of a gravitational field and a corresponding
acceleration of the reference system.\rq\rq

\bigskip

\subsection{The Role of Special Relativity.}

So far I did discuss Einstein's first principle of
equivalence in terms of Newton's theory of gravitation. Einstein had
set himself the task of studying how the principles of special
relativity from his 1905
paper\footnote{\foreignlanguage{ngerman}{\ }\foreignlanguage{ngerman}{Einstein,
Albert. }Ref. 1, p. 140.} \textit{On the Electrodynamics of Moving
Bodies} would affect Newton's theory of gravitation. Looking back from
a century later, we say that special relativity is based on the
representations of the Poincar\'e group of space{}-time translations
and Lorentz transformations while Newton's theory was based on those of
the Galilei group. This means that Newton's theory was simply
incompatible with special relativity.

Einstein's principle of equivalence gave him the clue to
search for a theory of gravitation based on special relativity. If
gravitation was locally nothing but a description of space and time
from an accelerated reference frame, he could succeed by studying
accelerated reference frames in special relativity. And this is what he
did. His deep physical intuition led him to two crucial conclusions:

Two identical clocks at rest in a gravitational field will
show a relative difference $\Delta \nu / \nu$ in their rate
$\nu$ given by $\Delta\Phi/c^2$, their
difference $\Delta\Phi$ in gravitational potential $\Phi$ divided
by the square of the velocity $c$ for light in a vacuum. The clock on the
higher potential, e.g., in the earth field at higher elevation, would
run slightly faster. This was first demonstrated in a terrestrial
experiment\footnote{\ Pound, R. V. and Rebka, G.A. ``Apparent Weight of
Photons'', Phys. Rev. Lett. \textbf{4}, 337{}-341 (1960).} by Robert
Pound and Glen Rebka in 1960, five years after Einstein's death. He had
not foreseen an experiment on earth as a test where for a difference in
height of ten meters $\Delta \nu / \nu$ is
$10^{-15}$, a millionth of a billionth. Einstein
wrote\footnote{\ Einstein, Albert. Ref. 1, p. 307. Einstein's solar
redshift is $2\times10^{-6}$, not, as translated here,
\lq\lq one part in two millionth\rq\rq.}

\lq\lq There exist \lq clocks\rq\ that are present at locations of
different gravitational potentials and whose rates can be controlled
with great precision; these are the producers of spectral lines. It can
be concluded from the aforesaid that the wavelength of light coming
from the sun's surface, which originates from such a producer is larger
by about one part in two million than that of light produced by the
same substance on earth.\rq\rq

It was only after 1960, as the conditions on the solar surface
were better understood, that Einstein's prediction for the sun could be
confirmed.

The other important result of his investigation concerned
the validity of his formula $ E = mc^{2} $. He stated at the
end of his
article\footnote{\foreignlanguage{ngerman}{\ }\foreignlanguage{ngerman}{Einstein,
Albert. Ref. 1, p. 311.}}:

\lq\lq Thus to each energy $E$ in the gravitational
field there corresponds an energy of position that equals the potential
energy of a \lq ponderable\rq\ mass of magnitude $ E/c^{2}$.
Thus the proposition, that to an amount of energy $E$ there
corresponds a mass of magnitude $E/c^{2}$, holds not only
for the \textit{inertial} but also for the \textit{gravitational} mass,
if the assumption introduced in Section 17 [the first principle of
equivalence] is correct.\rq\rq

Einstein also pointed out that in a gravitational field light
rays, not in the direction of the acceleration, would be bent. But the
correct formula for this process was not yet attained.

\subsection{A Forgotten Berichtigung}

Since explanations of Einstein's equivalence
principle are usually given without the use of special relativity, a
crucial detail of its formulation remains often unmentioned. In
Newtonian Theory this principle is true for extended bodies moving with
arbitrary velocities since the gravitational acceleration in this
theory is independent of velocity. In Relativity Theory this is no
longer the case. Here acceleration of a particle is a vector that is
always orthogonal to the tangent 4-vector of its world line. The
notion of {\it relative} acceleration exists only for particles
whose four-velocities agree. Interpreting the gravitational
acceleration of a falling object as minus the acceleration of the
reference system had to be restricted to objects at rest. When Einstein
wrote his Jahrbuch article in 1907 he was apparently not aware of this
limitation. It was a letter from Max Planck that had alerted him to
this fact when he published a \textit{Berichtigung} (an Erratum) in the
1908
Jahrbuch\footnote{\foreignlanguage{ngerman}{\ }\foreignlanguage{ngerman}{Einstein,
Albert Ref. 1. p. 317.}}.

The two systems $\Sigma_{1}$ and
$\Sigma_{2}$ of his Jahrbuch article, one accelerated and
the other in a gravitational field, must, therefore, not be considered
in motion with respect to each other. One has to think of the two
systems as one and the same system. There is no Poincar\'e
transformation between $\Sigma_{1}$ and
$\Sigma_{2}$ different from the identity. What
distinguishes the two systems is their different dynamical
interpretation: the question whether Einstein's system ${\Sigma}$ is
accelerated or suspended in a gravitational field. There are also
relativistic problems for extended bodies since a homogeneous
gravitational field in Minkowski's space-time modifies the
translation group. Relativity theory thus restricts the validity of the
Newtonian principle of equivalence to a local space-time event or to
a single world line. This was acknowledged by Einstein in the
{\it Erratum}. His considerations could only work for small
velocities, for small accelerations, only in a small neighborhood of an
event in space. He did not have the necessary concepts and mathematical
theories available to discover the relativistic gravitational field.
Even Minkowski's beautiful space{}-time picture that would have been of
help was demonstrated only later that year. The recognition that his
brilliant idea did not point the way to an extended relativistic
gravitational field must have been devastating. It would have
discouraged him to apply for the Beneke Prize if he had ever considered
it.

Ten years after his Yearbook article Einstein described his
first equivalence principle in loving detail in an account of the new
theory of gravitation in his book\footnote{\ Einstein, Albert.
\lq\lq Relativity\rq\rq. Penguin Books. NYC, NY 2008. p. 63.}
{\it Relativity, the Special and the General Theory}. This
description became the archetype of the Einstein elevator. The elevator
car, called {\it Kasten} by Einstein, was Englished into a
{\it chest} by his authorized translator Robert W. Lawson, a British
physicist who had studied German as a prisoner of war in Austria.
Einstein allowed his chest be pulled with constant acceleration to
reach arbitrary high velocities without mentioning the limitations of
Special Relativity. He apparently tried to erase the
{\it Berichtigung} from his memory. As far as I know, Einstein never
referred to his \textit{Erratum }again. Nor do his biographers.

\section{The Relativistic Gravitational Field}

\subsection{Twenty Years Later}

In May 1928 Einstein was bedfast with pericarditis. He wrote
to his friend Heinrich Zangger in Z\"urich\footnote{\ Einstein, Albert.
\lq\lq Einstein Archives\rq\rq, The Hebrew University of Jerusalem (EA), call
no. 40--69. \foreignlanguage{ngerman}{Reproduced in \lq\lq Albrecht
F\"olsing, }\foreignlanguage{ngerman}{\textit{Albert
Einstein}}\foreignlanguage{ngerman}{. Suhrkamp Verlag Frankfurt am Main
1994. Third Edition. p. 684. }}: \lq\lq In the tranquility of illness I have
laid a wonderful egg in the domain of General Relativity. Whether the
bird hatching from it will be vigorous and long-lived lies on the
knees of the gods. Meanwhile I approve the illness that so has blessed
me.\rq\rq\ On July 10, 1928 the bird appeared as \textit{Riemann-geometry
with keeping the Notion of Tele-parallelism}. In the introduction to
his paper\footnote{\foreignlanguage{ngerman}{\ }\foreignlanguage{ngerman}{Einstein,
A. ``Riemann{}-Geometrie mit Aufrecterhaltung des Begriffes des
Fernparallelismus''.
}\foreignlanguage{ngerman}{\textit{Sitzungsberichte der Preu{\ss}ischen
Akademie der Wissenschaften Phys.-Math. Klasse.
1928.XVII}}\foreignlanguage{ngerman}{. p.~217.}} he explained:

\lq\lq Characteristic for Riemann's geometry is that the
infinitesimal neighborhood of every point P has a Euclidean metric and
that the length of two line-elements belonging to the infinitesimal
neighborhoods of two points P and Q at a finite distance from each
other can be compared to each other. However, the notion of parallelism
of two such line-elements is missing. The notion of direction does
not exist for the finite. The theory to be proposed in the following is
characterized by the fact that it introduces for the finite besides the
Riemannian metric the \lq direction\rq, or rather the equality of direction,
or the parallelism.\rq\rq

\subsection{Ricci's Grid}

In his paper Einstein defined parallelism by using Ricci's
\textit{ennuples} that he now named {\it $n$-Beine}\,($n$-legs). He
called two vectors parallel if their frame components were
proportional. In a Euclidean space with frames based on ortho-normal
Cartesian coordinates this would result in Euclid's definition of
parallelism. However, based on Ricci's ennuples it extended the notion.
In particular, this meant now that parallel vectors of the same length
had identical frame components everywhere.

It was the great Italian geometer Gregorio
Ricci-Curbastro who generalized the concept of 3-dimensional frames
{\bf i}, {\bf j}, {\bf k} in the Euclidean space to such ortho-normal frames in
n-dimensional Riemannian manifolds. He introduced them in his 1895
paper\footnote{\ Ricci{}-Curbastro, Gregorio, \lq\lq Sulla Teoria degli
Iperspazi\rq\rq\ {\it Rend. Acc. Lincei Serie IV} (1895). p. 232--237.
Reprinted in : Gregorio Ricci{}-Curbastro, Opere, vol. I. Edizioni
Cremonense Roma 1956. p. 431--437.} {\it On the Theory of
Hyperspaces} and called them later \lq\lq ennuples\rq\rq. Ricci's ennuples
formed a frame of $n$ unit vectors ${\bold h}_{j}$
with $j = 1, {\dots}, n$. An arbitrary vector field
${\bold a}(x^{\lambda})$ at a point with coordinates
$x^{\lambda}$ would then be represented by its frame
components $a_{j}(x^{\lambda})$ in terms of
the unit vectors ${\bold h}_{j}(x^{\lambda})$ as
\begin{equation}
{\bold a} \, = \, a_{j} \, {\bold h}_{j}
\, \equiv \, \Sigma_{j = 1{\dots}n} \, a_{j} \, {\bold h}_{j} \, .
\end{equation}
The ${\bold h}_{j}$ were subject to the ortho-normality
conditions
\begin{equation}
{\bold h}_{j} \cdot {\bold h}_{k} \, = \, \delta_{jk} \, , \quad j, k = 1,2,{\dots},n
\end{equation}
with constants $\delta_{jk}$ vanishing for $j \neq k$ and
equal to one for $j = k$.

Vector fields can be visualized as stream-lines of a
stationary flow or as Faraday's lines of force. A non-vanishing
vector field in space generates a space-filling system of lines
through each point, known to mathematicians as a {\it congruence}.
Ricci's vision\footnote{\ Ricci{}-Curbastro, Gregorio ``Dei sistemi di
congruenze ortogonali i una variet\`a\par  qualunque'' \ \textit{Mem.
Acc. }\foreignlanguage{ngerman}{\textit{Lincei Serie 5 vol II
}}\foreignlanguage{ngerman}{(1896). p. 276--322. }Reprinted in:
Gregorio Ricci-Curbastro Opere Vol II Editore Cremonense. Roma 1957.
\ p. 1--61 } filled Riemann's $n$-dimensional space with $n$ congruences
orthogonal to each other creating a framework for the physical
components of vectors and tensors. This scaffold is not tied to the
coordinates. It serves as a reference body for measurements in
Riemann's manifold with Ricci's ennuples as tangent unit vectors in
each point providing the local scale along each line. I call this
reference body \lq\lq Ricci's Grid\rq\rq.

\subsection{Einstein Discovers Ricci's Grid}

In 1901 Ricci published, together with his student Tullio
Levi{}-Civita, a review paper\footnote{\ Ricci, Gregorio, and
Levi{}-Civita, Tullio. ``M\'ethodes de calcul diff\'erentielle absolu
et leurs applications.''
\foreignlanguage{ngerman}{\textit{Mathematische
Annalen}}\foreignlanguage{ngerman}{ 54 (1901) p. 125--201.}} of his
ingenious system of $n$-dimensional tensor analysis with its clever use
of upper and lower indices. This paper, written in French by two
Italians for a German journal, became the source from which Einstein,
helped by Marcel Gro{\ss}mann, derived the formal tools for his theory
of gravitation. Einstein did not find it easy learning the new
formalism. His friend Louis
Kollros\footnote{\foreignlanguage{ngerman}{\ }\foreignlanguage{ngerman}{Kollros,
L. \lq\lq Albert Einstein en Suisse Souvenirs\rq\rq\ in
}\foreignlanguage{ngerman}{\textit{F\"unfzig Jahre
Relativit\"atstheorie}}\foreignlanguage{ngerman}{ Helvetica Physica
Acta Supplementum IV 1956.}} remembered from that time Einstein's
scream \lq\lq Gro{\ss}mann, you've got to help me or I'm going nuts!\rq\rq

Einstein wrote to Arnold Sommerfeld\footnote{\ Einstein,
Albert. \lq\lq The Collected Papers of Albert Einstein\rq\rq\ volume 5. English
Translation. Anna Beck, Translator. Don Howard, Consultant. Princeton
UP Princeton NJ 1995. p.324.}:

\lq\lq I am now working exclusively on the gravitation problem
and believe that I can overcome all difficulties with a mathematical
friend of mine here. But one thing is certain: never before in my life
have I troubled myself over anything so much, and I have gained
enormous respect for mathematics, whose more subtle parts I considered
until now, in my ignorance, as pure luxury! Compared with this problem,
the original theory of relativity is child's play\rq\rq.

At first, Einstein did not use the full advantage of the formalism and
seems to have confined his study to the first chapter of the tensor
bible that did not use Ricci's ennuples and always worked with
coordinate components of vectors and tensors. It was apparently only in
1928 when he discovered his new geometry based on Ricci's ennuples that
Einstein apparently had progressed to the second chapter bearing the
title {\it La G\'eom\'etrie intrinseque comme instrument de calcul}.

Before discussing Einstein's new geometry it will be useful to
look at two examples, first its simplest derived from polar coordinates
in the Euclidean plane.

\subsection{Plane Polar Coordinates}

The metric of the Euclidean plane written in polar
coordinates $r$ and $\theta$ is given by
\begin{equation}
ds^{2} \, = \, dr^{2} \, + \, r^{2} \, d\theta^{2} \, .
\end{equation}
Since the lines $r = const$ and $\theta = const$ intersect under
right angles the frame of their unit tangent vectors ${\bold h}_{1}$ and ${\bold h}_{2}$
can be chosen as:
\begin{equation}
{\bold h}_{1} \, = \, \partial / \partial{\bold r} \, ,
\quad {\bold h}_{2} \, =
\, r^{-1} \partial / \partial \theta \, .
\end{equation}

In Einstein's geometry the auto-parallel lines are those
whose tangent vector is always parallel to itself. Those are the
{\it straightest} lines in his geometry. For the frames adapted to
plane polar coordinates such lines will intersect the lines $\theta = const$
under a fixed constant angle. Their equation is given by
\begin{equation}
r \, = \, a \, e^{\beta \theta} , \quad a > 0 \, , \quad \beta \neq  0 \, ,
\end{equation}
with real constants $a$ and $\beta$. They are known as logarithmic
spirals. The coordinate lines themselves are their degenerate cases

A remarkable feature for rectangles of straightest lines in
Einstein's geometry is the following: a rectangle formed by the two
auto-parallels $ r = r_{1} $ and $ r = r_{2} $
and the other pair of auto-parallels $ \theta = \theta_{1} $ and $ \theta = \theta_{2} $
will have one set of opposite sides with equal length $ r_{2} - r_{1} $
and the other set with unequal lengths
$ r_{2} ( \theta_{2} - \theta_{1}) $ and $ r_{1} ( \theta_{2} - \theta_{1} ) $,
respectively (Fig. 1).

\subsection{Einstein's Chest in Minkowski's Space-Time}

Einstein's chest was accelerated in the direction
of the $x$-axis with constant acceleration $\gamma$. Such a motion was
discussed by
Max Born\footnote{\foreignlanguage{ngerman}{\ }\foreignlanguage{ngerman}{Born,
M. Annalen der Physik, Leipzig.} Volume 30 (1909). p. 1.} in 1909. For
the description of this system I can use Ricci's ennuples in Minkowski
space-time with metric
\begin{equation}
ds^{2} \, = \, dx^{2} \, + \, dy^{2} \, + \, dz^{2} \, + \, (icdt)^{2} \, .
\end{equation}
A point on the floor of the chest describes, due to its
constant acceleration, a world line that is an equilateral hyperbola
with constant R.
\begin{equation}
(x/R)^{2} \, + \, (ict/R)^{2} \, = \, 1 \, , \quad x > 0 \, .
\end{equation}
The acceleration, the inverse radius of this pseudo-circle,
is $ \gamma \, = \, c^{2}/R $
where $c$ is the speed of light. For $ \gamma = 9.8 \, m s^{-2} $, $R$ equals one light-year.
If $h$ is the height of the chest, the equation for a point on its ceiling is
\begin{equation}
\left(\frac{x}{R+h}\right)^{2} \, + \, \left(\frac{ict}{R+h}\right)^{2} \, = 1 \, , \quad
x > 0 \, .
\end{equation}
Its acceleration, given by $ \gamma = c^{2} /(R+h) $, is
just a tiny bit smaller. For the gravitational picture it is convenient
to introduce coordinates in the $x$-$ct$-plane of Minkowski's
space-time by the analogous transition from Cartesian to polar
coordinates
\begin{equation}
x \ = \ (R + \xi) \, Ch \, \tau \, , \quad
ct \, = \, (R + \xi) \, Sh \, \tau \, .
\end{equation}
The Minkowski metric takes then the form
\begin{equation}
ds^{2} \, = \,
d\xi^{2} \, + \, dy^{2} \, + \, dz^{2} \, - \, (R + \xi)^{2} \, d\tau^{2} \, .
\end{equation}
While the hyperbolae $ \xi = const $ in the
$\xi$-$\tau$-plane are the analogs of concentric circles in the
Euclidean polar coordinates, the lines of equal time $\tau$ correspond
to the radii of those circles. This coordinate system is orthogonal in
the space-time metric. In the new coordinates the frame vectors are
drawn along the coordinate lines.
The frame vectors $ {\bold h}_{j} $ are in the chest
\begin{eqnarray}
{\bold h}_{1} \, &= \, \partial / \partial \xi \, , \quad 
{\bold h}_{2} \, &= \, \partial / \partial y \, , \quad \nonumber \\
{\bold h}_{3} \, &= \, \partial / \partial z \, , \quad
{\bold h}_{4} \, &= \, i (R + \xi)^{-1}
\partial / \partial \tau \, .
\end{eqnarray}
The fourth vector, tangent to the hyperbolae, is an imaginary one. The
acceleration in the chest at a height $\xi$ above the floor is given
by $ \gamma = c^{2}/(R + \xi) $ The radius of
curvature of the hyperbolic world line of any fixed point in the chest
shows the state of acceleration in Einstein's system
$ \Sigma_{1} $ (Fig. 2).

\subsection{The Chest Viewed as a Gravitational Field}

However, I can also view the chest as system
$\Sigma_{2}$ by using Einstein's geometry. Then fixed
points in the chest proceed on auto-parallel world lines with tangent
unit 4-vector ${\bold h}_{4}$ defining a state of rest in
the chest. If an object is dropped in the chest it will, neglecting air
resistance, describe a geodesic world line. If the object is released
at time $ \tau = 0 $ at height $\xi_{0}$ above the
floor its motion is described by the equations
\begin{equation}
R + \xi_{0 } \, = \, (R + \xi) \, Ch \, \tau \, , \quad
ct = (R + \xi) \, Sh \, \tau \, ,
\end{equation}
with $ (y, z) = const $.
This gives for small times $t$
\begin{equation}
\xi \, = \, \xi_{0} \, - \, \frac{1}{2} \, \frac{c^{2}}{R + \xi} \, t^{2} \, ,
\end{equation}
leading, at $t = 0$, to an acceleration
$ \gamma \, = \, - c^{2}/(R + \xi_{0})$.
The object dropped at $t = 0$ proceeds on the geodesic
$x = R + \xi_{0} = const$. At $t = 0$ it's 4-velocity
${\bold u}$, tangent to the geodesic, is
equal to the frame vector
${\bold h}_{4}$. At that moment its gravitational
acceleration equals minus the acceleration of the system
$\Sigma_{1}$. The gravitational field is not exactly
homogeneous. There exist homogeneous gravitational fields in Minkowski
space-time with constant acceleration\footnote{\ Schucking, Engelbert
and Surowitz, Eugene. \lq\lq Einstein Fields\rq\rq. A book
manuscript, unpublished.}. I have chosen here the case of a static
field because it is described by the {\it Rindler coordinates} that
go back to Levi-Civita\footnote{\ Levi-Civita, Tullio, \lq\lq Statica
Einsteiniana,\rq\rq\ Rendiconti della R. Accademia dei Lincei, ser. 5,
1\textsuperscript{st} sem., vol. 26 (1917). p. 458.}.

\subsection{The Gravitational Red-shift}

If a light wave is emitted vertically upwards from the floor
of Einstein's chest, its frequency $\nu$ will be red-shifted when
received at the ceiling of the chest. It is sometimes claimed that the
existence of such a gravitational red-shift forces us to admit a
non-vanishing space-time curvature\footnote{\ Schild, Alfred.
Lectures in Applied Mathematics, Vol. 8. American Mathematical Society,
Providence R.I. (1967).}. These arguments are not convincing since we
are dealing with flat Minkowski space-time. What the arguments
actually get at is the Einstein geometry of Minkowski space-time when
interpreted as a static gravitational field.

The existence of the red-shift is immediately clear in the
acceleration picture $\Sigma_{1}$ where source and receiver
of the light wave are accelerated. If the wave is emitted upwards from
the floor, then by the time of reception the receiver at the ceiling
will have acquired a speed $V$ of recession of about
$ V = \gamma h/c $ if $V << c$.
The red-shift appears then as the Doppler shift
\begin{equation}
- \, \frac{\delta \nu}{\nu}
\, = \, \frac{V}{c}
\, = \, \gamma \, \frac{h}{c^{2}} \ = \ \frac{\Delta\lambda}{\lambda} \, .
\end{equation}
The gravitational red-shift has a simple explanation in Einstein's
geometry (Fig. 3).
One can refer to the discussion of polar coordinates on the
Euclidean plane. Drawing in the $\xi$-$\tau$-plane the
rectangle bounded on bottom and top by the auto-parallel lines
$ \tau = 0 $ and $ \tau = 1 $, respectively, and adding its left and
right sides by the auto-parallel lines $ \xi = 0 $ and $ \xi = h $,
respectively, one notices that opposite sides of the rectangle are
parallel. Top and bottom of the rectangle have both length $h$. The left
side of the rectangle has {\it length} $R$, i.e., proper time $R/c$,
while the right side has {\it length} $R + h$, i.e., proper time 
$(R + h)/c$. This missing length $h$, or missing proper time $h/c$, is the
geometrical description of the gravitational red-shift in Einstein's
geometry. The relative gap $h/R$ equals the relative change in wavelength
${\Delta}$${\lambda}$/${\lambda}$. It says in physical terms that a
standard clock at the ceiling of Einstein's chest runs faster than a
standard clock on the floor. Such non-closure of rectangles, or
parallelograms, is the typical feature of Einstein's geometry. It has
nothing to do with the curvature of space-time since the Minkowski
plane has zero curvature.

The analogy with the Einstein geometry in the Euclidean plane based on
polar coordinates goes even further than expected. If one asks for the
auto-parallel lines of constant velocity $\boldsymbol\beta$ in the vertical
direction one is looking for the analogs of the logarithmic spirals in
the Euclidean plane. Their equation
turns out to be the same, except
for new names of the variables
\begin{equation}
R \, + \, \xi \, = \, a \, e^{\beta \tau} \, .
\end{equation}

Jakob Bernoulli was the first using polar coordinates systematically and
studied the properties of the logarithmic spiral. He became so fond of
it he willed this curve engraved on his tombstone at the M\"unster in
Basel, Switzerland, where it can still be seen today. It carries the
inscription {\it eadem mutata resurgo} (Though changed I rise
again). Unfortunately, the stone mason made it an Archimedean spiral.

\subsection{What Is a Gravitational Field?}

The brief answer is:
A GRAVITATIONAL FIELD IS A TELE-PARALLEL RICCI GRID.

Its mathematical characterization is quite simple: the
commutator of the tele{}-parallel vector fields of the Grid
\begin{equation}
\left[ \, {\bold h}_{j}, \, {\bold h}_{k} \, \right]
\, = \, - \, {\bold h}_{l} \, T^{l}{}_{jk} \, , \quad
T^{l}{}_{jk} \, = \, - \, T^{l}{}_{kj} \, ,
\end{equation}
gives rise to the torsion tensor $T^{l}{}_{jk}$
that is skew-symmetric in its lower indices.

(If we use an imaginary time-like vector we need not distinguish
between upper and lower indices.) This three-index tensor describes
the field strength of the gravitational field. For the example in
equation (11) we obtain for the only component of the torsion tensor
different from zero
\begin{equation}
T^{4}{}_{14} \, = \, - \, T^{4}{}_{41} \, = \, 1/(R + \xi) \, .
\end{equation}

One must wonder why Einstein did not recognize that his new
geometry of the tele-parallel Ricci Grid ended the long-sought
search for the tensorial description of his relativistic gravitational
field. Here was the precise definition of the {\it reference body}
necessary for measurements and the elusive reference
mollusk\footnote{\ Ref. 13, p. 112.}. The Ricci Grid was just a
mathematical model for the matter serving as measuring instrument for
the field.

I can only speculate why this wasn't obvious when he
introduced his new geometry for Riemannian manifolds:

Since the beginning of the 1920's Einstein had largely
dropped research on his theory of gravitation though he still
supervised work of collaborators on, e.g., the theory of motion in
General Relativity. Instead, Einstein had become obsessed with finding
a home for electromagnetism in a geometrical theory that included
gravitation. This question occupied him for the last third of his
life\footnote{\ Goenner, Hubert \lq\lq On the History of Unified Field
Theories\rq\rq\ {\it Living Reviews in Relativity}
http://www.livingreviews.org/lrr-2004-2}.
When he discovered {\it Fernparallelismus}\footnote{\ Sauer,
Tilman \lq\lq Field equations in teleparallel spacetime: Einstein's
{\it Fernparallelismus} approach towards unified field theory\rq\rq.
arXiv:physics/0405142v1 26 May 2004 and HISTORIA MATHEMATICA
[doi101016/jhm2005.11.005] .}
\textit{ }\ in 1928 he wanted to use its 4-dimensional geometry for a
field theory combining electromagnetism with gravitation. However, he
saw in the frame vectors not the description of the reference body but
the manifestations of E\&M. Einstein believed that the tensor $T^{l}{}_{jk}$
contained besides the degrees
of freedom of the gravitational field also those of electromagnetism,
e.g., initially he identified a contraction of the tensor with a
multiple of the electromagnetic four-potential. It was not clear to
him that the tensor $T^{l}{}_{jk}$ was
simply equivalent to the tensor of accelerations.

\section{The First Principle of Equivalence, Final Version}
\subsection{The Susskind Principle of Equivalence}

\lq\lq THE EQUIVALENCE PRINCIPLE: GRAVITY IS INDISTINGUISHABLE
FROM ACCELERATION.\rq\rq\footnote{\ Susskind, Leonard. \lq\lq The Cosmic Landscape\rq\rq.
Little, Brown and Co. New York, NY 2005 p.~347.}

This brief formulation expressed, perhaps, Einstein's aim to
generalize his first principle of equivalence that he had formulated
for homogeneous gravitational fields in 1907. Since we now have a
precise invariant definition of a gravitational field, we can
investigate whether Susskind's formulation can be justified. For that
purpose we need a definition of acceleration. It is sufficient to have
such an expression for the basis vectors ${\bold h}_{j}$.
It is given by Ricci's covariant derivation of the vector
${\bold h}_{j}$ in direction ${\bold h}_{k}$ written as
\begin{equation}
\nabla_{{\bold h}_k} {\bold h}_{j} \, = \, - \, \gamma^{l}{}_{jk} \, {\bold h}_{l} \, ,
\end{equation}
{\selectlanguage{english}
where the $\gamma_{jkl}$ are known as Ricci's
rotation coefficients. In his brief 1895 paper ``Sulla Teoria Degli
Iperspazi'' that introduced the coefficients, Ricci showed that they
were skew{}-symmetric in their first pair of
indices\footnote{\foreignlanguage{ngerman}{\ }\foreignlanguage{ngerman}{Ricci{}-Curbastro,
Gregorio Ref. 17.}}}
\begin{equation}
\gamma_{jkl} \, = \, - \, \gamma_{k j l} \, .
\end{equation}
Ricci's covariant derivative has vanishing torsion. That is expressed by
the equation
\begin{equation}
\nabla_{{\bold h}_{k}} {\bold h}_{j} \, - \, \nabla_{{\bold h}_{j}} {\bold h}_{k}
\, - \, \left[ \, {\bold h}_{k} , \, {\bold h}_{j}
\, \right] \, = \, 0 \, .
\end{equation}
Using equation (16) gives with equations (18) and (20) the \newline
FINAL VERSION OF THE EQUIVALENCE PRICIPLE:
\begin{equation}
- \, \gamma^{l}{}_{jk}\, + \, \gamma^{l}{}_{kj} \, = \, T^{l}{}_{jk} \, .
\end{equation}
The left-hand side of this equation is given by the acceleration of
the Ricci Grid, while the right-hand side carries the tensor of the
gravitational field strength. The right-hand side determines also the
Ricci rotation coefficients that form the contorsion tensor. Using
equation (19) one easily derives from (21) that
\begin{equation}
\gamma_{ljk} \, = \,
\frac{1}{2} \left( \, - \, T_{ljk} \, + \, T_{klj} \, - \, T_{jkl}
\, \right) \, .
\end{equation}
This equation confirms that we have a true equivalence of gravitation
and acceleration.

\subsection{Ricci's Coefficients in Space-Time}

A Ricci Grid at an event in space{}-time is a mathematical
model for an infinitesimal rigid body at rest defining units of length
and time. A different ennuple at the same event is obtained through a
Lorentz transformation of the ennuple. The six parameters of that
transformation can be given by an angular orientation vector $\boldsymbol \phi$
and a velocity vector ${\bold v}$. A neighboring
ennuple differs by an infinitesimal Lorentz transformation described by
the skew-symmetric first pair of indices in Ricci's $\gamma$-tensor.
The third index of the tensor gives the gradient of the infinitesimal
Lorentz transformation.

If indices $a, b, c$ run through the numbers from 1 to 3,
$ - \, i \, \gamma_{ab4} $ describe angular velocity and
$ - \, \gamma_{4a4} $ acceleration against absolute
space for the infinitesimal rigid body.

Further, $ \gamma_{abc} $ are the spatial
gradients of the orientation vector $\boldsymbol \phi$ for the rigid
body, while $ - \, i \, \gamma_{4ab} $ is the spatial
gradient of the vector \textbf{v}.
Altogether, one can say that the coefficients are a generalization of the
Pauli-Lubanski vector for the vectors in Ricci's ennuples. Through the
split into space and time one describes all 24 components of what we
call {\it acceleration}.

\section{Some History of the Tele-Parallel Ricci Grid}

\subsection{Hessenberg}

When Einstein discovered tele-parallelism in 1928
he was apparently not aware of the fact that the tele-parallel Ricci
Grid had been discovered a dozen years earlier. The man who first
recognized this possibility was a professor of mathematics in Breslau,
well known in the profession for his work on the foundations of
geometry. In a paper finished in June 1916 Gerhard
Hessenberg\footnote{\foreignlanguage{ngerman}{\ }\foreignlanguage{ngerman}{Hessenberg,
Gerhard, \lq\lq Vektorielle Begr\"undung der Differentialgeometrie\rq\rq.
}Mathematische Annalen vol. 78, p. 187--217 (1917).} replaced
Christoffel's cumbersome calculations by an invariant co-vector
method that reached later an even more elegant form in \'Elie Cartan's
papers\footnote{\ Cartan, \'E. \lq\lq Riemannian Geometry in an Orthogonal
Frame\rq\rq. World Scientific. \foreignlanguage{ngerman}{Singapore 2001.}}.
Hessenberg discovered the torsion tensor $T_{kjl}$ and
the contortion tensor. By introducing auto-parallel curves for a
tele-parallel Ricci Grid he proved that the torsion tensor vanishes
if and only if all auto-parallel curves are geodesics, that is,
shortest lines in the Riemann metric. Hessenberg was the first to
discover that the geometry of a gravitational field is characterized by
the torsion of teleparallelism.

{\selectlanguage{english}
\ I have nowhere seen his discovery of this special kind of torsion
acknowledged. Hermann
Weyl\footnote{\foreignlanguage{ngerman}{\ }\foreignlanguage{ngerman}{Weyl,
H. \lq\lq Raum . Zeit . Materie\rq\rq. Springer Verlag
7}\foreignlanguage{ngerman}{\textsuperscript{th}}\foreignlanguage{ngerman}{
ed. 1988. p.332.}}, who missed finding torsion when generalizing the
notion of connections, refers to Hessenberg only by crediting him with
the proof that the symmetry of the Riemann tensor in its first and
second pairs of the indices follows from the cyclic symmetry in the
last three indices. Appendix A offers a few notes about Hessenberg's
1916 paper.}

\subsection{Cartan}

Torsion was named by \'Elie Cartan\footnote{\ Cartan, \'E.
Comptes Rendus Acad. Sci.1922. vol. 174, p. 593.} and announced in a 3
page note in Comptes Rendus in March 1922 with the title
{\it A Generalization of the Riemann Curvature and the Spaces with Torsion}.
Einstein learned about this new revolutionary concept in geometry only
four weeks later. His friend Paul Langevin had invited him to lecture
at the Coll\`ege de France in Paris on March 31, 1922.

In the aftermath of World War I, the first lecture by a
professor from the country of the archenemy was a highly charged
political affair. To cut down on demonstrations, it was by invitation
only, and the French Prime Minister Paul Painlev\'e stood at the door
checking. During this lecture week, Jacques Hadamard, professor at the
Coll\`ege de France gave a party for Einstein. Among his guests, who
was to meet Einstein there, was \'Elie Cartan, the world authority on
Lie algebras and the greatest geometer of his time. Cartan thought
torsion might have important physical applications and used the
occasion to tell Einstein about his recent discovery. He tried to
explain the novel concept to him by the
example well known to map makers and navigators of tele-parallelism
arising from polar coordinates on the sphere. But apparently neither
Cartan nor Einstein \ realized that this example held the key to
Einstein's first equivalence principle.
Einstein\footnote{\foreignlanguage{ngerman}{\ }\foreignlanguage{ngerman}{Debever,
Robert (ed.) \lq\lq \'Elie Cartan {--} Albert Einstein. }Letters on
Absolute Parallelism 1929--1932\rq\rq.
Princeton University Press. 1979.} wrote to
Cartan 7 years later about his introduction to tele-parallelism:

\lq\lq I didn't at all understand the
explanations you gave me in Paris; still less was it clear to me how
they might be made useful for physical theory.\rq\rq

\subsection{Weitzenb\"ock}

The occasion of his
letter\footnote{\foreignlanguage{ngerman}{.Debever, Ref. 51. p.4}} was
that Einstein had attempted using tele-parallelism for a generalized
field theory of gravity and electromagnetism and Cartan had reminded
him that tele-parallelism was a special case of Cartan's torsion.
But since Einstein's first paper on torsion in 1928 he had learned that
Roland Weitzenb\"ock had also published papers on torsion. In fact, in
his paper \lq\lq Differential Invariants in Einstein's Theory of
Tele-parallelism\rq\rq\ Weitzenb\"ock\footnote{\foreignlanguage{ngerman}{\ }\foreignlanguage{ngerman}{Weitzenb\"ock,
R., Sitzungsberichte der Preu{\ss}ischen Akademie der Wissenschaften
phy.{}-math. Klasse (Berlin) 1928. p. 466.}} had given a supposedly
complete bibliography of papers on torsion without mentioning Elie
Cartan. A bizarre circumstance suggested that this omission may have
been deliberate. Weitzenb\"ock of Amsterdam University in the
Netherlands was an Austrian K. u. K. army officer before WWI. In 1923
he had published a modern monograph on the Theory of Invariants that
included Tensor
Calculus.\footnote{\foreignlanguage{ngerman}{\ }\foreignlanguage{ngerman}{Weitzenb\"ock,
R. \lq\lq Invariantentheorie\rq\rq. Nordhoff, Groningen 1923.}} In the innocent
looking Preface, one finds that the first letter of the first word in
the first 21 sentences spell out:

NIEDER MIT DEN FRANZOSEN (down with the French).

To set the record straight, Einstein invited
Cartan\footnote{\foreignlanguage{ngerman}{\ }\foreignlanguage{ngerman}{Cartan,
\'E. Annalen der Mathematik, vol.102 (1930) p. 698.}} to write about
the history of torsion in the Annalen der Mathematik.
Neither Cartan, nor Weitzenb\"ock in their history of torsion
mentioned Hessenberg who had died in 1925.

\section{What is General Relativity?}

\subsection{According to Synge and Bondi}

When after Einstein's death in 1955 a new generation
of differential geometers and mathematically oriented theoretical
physicists re-discovered General Relativity, the vague statements on
equality of all motions based on the first principle of equivalence had
lost their appeal. The Irish mathematician John Synge wrote in 1960 in
the Introduction to his book\footnote{\ Synge, John, \lq\lq Relativity: The
General Theory. Elsevier. New York, NY 1960. Preface,
p.\textsc{ix--x}.} \textit{Relativity, the General Theory}: \lq\lq {\dots}I
have never been able to understand this principle.\rq\rq\ And he went on:
\lq\lq Does this mean that the effects of a gravitational field are
indistinguishable from the effects of an observer's acceleration? If
so, it is false. In Einstein's theory, either there is a gravitational
field or there is none, according as the Riemann tensor does or does
not vanish. This is an absolute property; it has nothing to do with any
observer's worldline. Space-time is either flat or curved, and in
several places of the book I have been at considerable pains to
separate truly gravitational effects due to curvature of space-time
from those due to curvature of the observer's worldline (in most
ordinary cases the latter predominate). The Principle of Equivalence
performed the essential office of midwife at the birth of general
relativity, but, as Einstein remarked, the infant would never have got
beyond its long-clothes had it not been for Minkowski's concept. I
suggest that the midwife be now buried with appropriate honours and the
facts of absolute spacetime be faced.\rq\rq

At Einstein's Centenary in 1979 Hermann Bondi celebrated him
with the essay \footnote{\ Bondi, Hermann in \lq\lq Relativity, Quanta, and
Cosmology\rq\rq\ edited by Francesco de Finis. Volume1. Johnson Reprint
Corporation. New York, NY 1979. p.181.} 
{\it Is \lq\lq General Relativity\rq\rq\ Necessary for
Einstein's Theory of Gravitation?} Bondi wrote: \lq\lq From this point of
view, Einstein's elevators have nothing to do with gravitation, they
simply analyse inertia in a perfectly Newtonian way. Thus the notion of
general relativity does not in fact introduce any post-Newtonian
physics; it simply deals with coordinate transformations. Such a
formalism may have some convenience, but physically it is wholly
irrelevant. It is perhaps rather late to change the name of Einstein's
theory of gravitation, but general relativity is a physically
meaningless phrase that can only be viewed as a historical memento of a
curious philosophical observation.\rq\rq

In the unsuccessful attempt of finding a
mathematical formulation for General Relativity the field of research
narrowed into \lq\lq Einstein's Theory of Gravitation\rq\rq. But Einstein's
Theory of Gravitation was, simply speaking, \lq\lq Einstein's Field
Equations\rq\rq. What then was General Relativity built on the principle of
equivalence?

\subsection{Einstein Vindicatus}

A gravitational field defined as a tele-parallel
Ricci Grid is, in general, a non-inertial system for a finite region
of the space-time manifold. It also defines a reference body in a
state of acceleration and rotation against absolute space-time.
However, such gravitational fields appear already in Minkowski
space-time.

We consider now generalized Lorentz transformations of the
Ricci Grid. By that we mean space-time dependent transformations
\begin{equation}
{\bold h}^\prime_{m} \, = \,
A_{m}{}^{j}(x^{\lambda}) \, {\bold h}_{j} \, , \quad
A_{mj} \, A^{m}{}_{k} \, = \, \delta_{jk} \, .
\end{equation}
Such an infinitesimal transformation is given by
\begin{equation}
A_{m}{}^{j} \, = \,
\delta_{m}{}^{j} \, + \, \varepsilon_{m}{}^{j} \, ,
\quad \varepsilon_{mk} \, = \, - \, \varepsilon_{km} \, ,
\end{equation}
where the $ \varepsilon_{mk} $ are infinitesimal skew-symmetric
functions of the
coordinates. Besides the tensor transformation of the gravitational
field strength $ T^{l}{}_{jk} $ the terms
\begin{equation}
\delta T^{l}{}_{jk} \, = \, d \, {\varepsilon}_{k}{}^{l} \left({\bold h}_{j} \right)
\, - \, d \, {\varepsilon}_{k}{}^{l}
\left( {\bold h}_{k} \right)
\end{equation}
now appear due to the space-time dependence of the Lorentz
transformations.
GENERAL RELATIVITY MEANS:
INVARIANCE UNDER GENERALIZED LORENTZ TRANSFORMATIONS.

As the Korean physicist
Y.M. Cho showed in his paper\footnote{\ Cho, Y. M., Physical Review D 14 (1976), 2521.}
{\it Einstein Lagrangian as the translational Yang-Mills
Lagrangian} that the Lagrangean density
\begin{equation}
{\pounds} \, = \,
\kappa^{-2} \left( - \, \det h^{\mu}{}_{j} \right)^{1/2}
\left[ \, 
\frac{1}{4} \, T_{ijk} \, T_{ijk}
\, + \, \frac{1}{2} \, T_{ijk} \, T_{ijk}
\, - \, T_{ijj} \, T_{ikk}
\, \right]
\end{equation}
giving the Einstein vacuum field equations is, in fact, invariant under
generalized Lorentz transformations (I have used here an imaginary
${\bold h}_{4}$). Einstein's General Relativity shows the
egregious {\it physical} property of being invariant against a
change of reference body.



\section{ APPENDIX A}
\subsection{Hessenberg's {\it Vectorial Foundation of Differential Geometry}}
\setcounter{equation}{0}

Hessenberg introduces an $n$-leg, ${\underline p}_{j}$
$( j, k = 1, {\dots}, n )$, into every point of a $n$-dimensional Riemannian
manifold. I simplify his representation by taking these $n$-legs to be
orthonormal. Then all indices can be kept downstairs. I refer to
equations in his paper by putting them into square brackets \lq\lq [..]\rq\rq.
This gives his equation [24]
\begin{equation}
{\underline p}_{j} \, \cdot \, {\underline p}_{k} \, = \, \delta_{jk} \, .
\end{equation}
He defines the differential one-form $ db_{jk} $ in [39] by 
(I lower his indices according to footnote on his page 198)
\begin{equation}
d{\underline p}_{j} \cdot {\underline p}_{k} \equiv \, db_{jk}
\end{equation}
skew-symmetric in indices $j$ and $k$. His equation [41] gives
\begin{equation}
db_{jk} \, + \, db_{kj} \, = \, 0 \, .
\end{equation}
Hessenberg calls the differential one-form $ db_{jk} $ the \lq\lq Orientation Tensor\rq\rq.
Nowadays one would write
\begin{equation}
d{\underline p}_{j} \, = \, \omega_{jl} \, {\underline p}_{l} \, , \quad
d{\underline p}_{j} \, \cdot \, {\underline p}_{k} \, = \, \omega_{jk} \, , \quad
\omega_{jk} \, + \, \omega_{kj} \, = \, 0
\quad\quad (2',3')
\nonumber
\end{equation}
where $\omega_{jk}$ is now the connection one-form.
We thus have to identify $ db_{jk} $ with $\omega_{jk}$.
He next introduces in [47] a cogredient differential $d{\underline A}$ of a tensor $A$
with $\alpha$ indices
\begin{equation}
{\underline A} \, = \,
A_{j_1 \dots j_\alpha} \, {\underline p}_{j_1} \dots {\underline p}_{j_\alpha}
\end{equation}
in terms of the covariant differentials $ \delta A_{j_{1} \dots j_{\alpha}} $
\begin{equation}
d{\underline A} \, = \,
\delta A_{j_1 \dots j_\alpha} \, {\underline p}_{j_1} \dots {\underline p}_{j_\alpha}
\end{equation}
where this differential is defined by [48]
\begin{equation}
\delta A_{j_1 \dots j_\alpha} \, = \,
dA_{j_1 \dots j_\alpha} \, - \, db_{j_1k} \, A_{k \dots j_\alpha}
\, - \, \dots \, - \, db_{j_{\alpha} k} A_{j_1 \dots k} \, .
\end{equation}
In section 20 on page 205 he introduces differential one-forms
$ \omega_{j} $ that give the Riemannian metric
\begin{equation}
ds^{2} \, = \, \omega_{j} \, \omega_{j} \, .
\end{equation}
These differential forms $\omega_{j}$ are his $ u^{j\rho} \, dt_{\rho} $
(and confusingly also denoted as $ du^{j}$).
They are dual to his orthonormal vectors $ {\underline p}_{k} $
\begin{equation}
\omega_{j} ( \, {\underline p}_{k}) \, = \, \delta_{jk} \, .
\end{equation}
Hessenberg's equation [87] is the necessary and sufficient condition
that all straightest lines are geodesics in his more general geometry.
In this case the connection form specializes to
$ \omega'_{jk} $ obeying what appears now as the first
Cartan structural equation for vanishing torsion
\begin{equation} 0
\ = \ - \, d\omega_{j} \, + \, \omega'_{jk} \wedge \omega_{k} \, .
\end{equation}
His notation \lq\lq $D_{12} \, u^{j}$\rq\rq\ 
precisely explained in the first footnote on page 211 with the opposite
sign against Cartan's \lq\lq $d$\,\rq\rq-operator. In this case \ Hessenberg's
one-form $db_{jk}$ specializes into the Levi-Civita
connection form $\omega'_{jk}$.
On the other hand Hessenberg's equation [94] reads now
\begin{equation}
- \, d\omega_{j} \, = \,
\frac{1}{2} \, U_{ljm} \, \omega_{l} \wedge \omega_{m} \, , \quad
U_{ljm} \, = \, - \, U_{mjl} \, .
\end{equation}
This equation according to Cartan's first
structural equation defines the right-hand side of (10) as the
negative of the torsion form $ \Theta_{j} $ for a vanishing
connection that defines the teleparallelism of Hessenberg's straightest
lines
\begin{equation}
\frac{1}{2} \, U_{ljm} \, \omega_{l} \wedge \omega_{m} \, = \,
- \, \Theta_{j} \, = \,
- \, d\omega_{j} \, + \, \omega_{jk} \, , \quad
\omega_{jk} \, = \, 0 \, .
\end{equation}
Comparing now equations (9) and (10) he needs
the development of the Levi-Civita connection form
$\omega'_{jl}$ in terms of the $\omega_{m}$.
This gives the Ricci rotation coefficients $h_{jlm}$
\begin{equation}
\omega'_{jl} \, = \, h_{jlm} \omega_{m} \, , \quad
h_{jlm} \, = \, - \, h_{ljm} \, .
\end{equation}
This is the meaning of [97] where the Christoffel symbol of the
second kind vanishes because of our simplification (1). \ From (9) and
(10) follows now
\begin{equation}
0 \, = \, \frac{1}{2} \, U_{ljm} \, \omega_{l} \wedge \omega_{m}
       \, + \, h_{ljm} \, \omega_{l} \wedge \omega_{m} \, ,
\end{equation}
or simply Hessenberg's equation [98]
\begin{equation}
0 \, = \, U_{ljm} \, + \, h_{ljm} \, - \, h_{mjl} \, . 
\end{equation}
By cyclic interchange of the indices one obtains the
equations
\begin{equation}
0 \, = \, U_{jml} \, + \, h_{jml} \, - \, h_{lmj}
\end{equation}
and
\begin{equation} 
0 \, = \, U_{mlj} \, + \, h_{mlj} \, - \, \ h_{jlm} \, . 
\end{equation}
Then (14) + (15) - (16) gives
\begin{equation}
2 \, h_{mjl} \, = \, U_{ljm} \, + \, U_{jml} \, - \, U_{mlj} \, .
\end{equation}
Changing to the indices used in
Hessenberg's second footnote on page 211 one obtains
(I replace the index ``i'' by ``j'' because Microsoft insists on capitalizing it
when it comes first)
\begin{equation}
2 \, h_{jlk} \, = \, U_{ljk} \, + \, U_{klj} \, - \, U_{jkl} \, .
\end{equation}
This does not agree with the
expression in Hessenberg's second footnote on page 211. The reason is
that he re-defines $U$ by turning its upper index into its first
lower index instead of into its second as he stated as a general rule
in the footnote on page 198. The interchange of the first two indices
in $U$ turns (18) into $U'$, skew-symmetric in its last two indices,
\begin{equation}
2 \, h_{jlk} \, = \, U'_{jlk} \, + \, U'_{lkj } \, - \, U'_{kjl} \, .
\end{equation}
This agrees with Hessenberg.
With $U'_{jlk}$ being the negative of the torsion tensor
the tensor $ h_{jlk} $ becomes now the negative of the contorsion tensor.
The contorsion tensor $ g_{lkj} $ becomes by cyclic permutation of the indices
\begin{equation}
2 \, g_{jlk} \, = \, C_{lkj} \, + \, C_{kjl } \, - \, C_{jlk} \, ,
\end{equation}
where the tensor $C_{lkj}$ is skew-symmetric in its first two indices.
With
\begin{equation}
C_{lkj} \, = \, - \, U'_{jlk} \, ,
\end{equation}
comparing
(19) and (20) gives
\begin{equation}
h_{jlk} \, = \, - \, g_{jlk} \, ,
\end{equation}
identifying the contortion tensor with the negative Ricci rotation
coefficients of Christoffel's covariant derivative (which became called
the Levi-Civita connection). In this way Hessenberg
discovered a special case of torsion, namely, as we would nowadays say,
a case where the symmetric part of the connection coefficients
vanishes, a case of tele-parallelism. The fact that he has a
geometric interpretation for it in terms of auto-parallel curves
shows that he is writing about a geometric phenomenon, the discrepancy
between the straightest and the shortest curves in a geometry with
torsion. This is exactly the example that Cartan used to explain
his geometry to Einstein pointing out the distinction between rhumbs
and geodesics on the sphere.


\bigskip
\centerline {\bf Coda}
I am grateful for exchange of views with Friedrich Hehl, Roger Penrose,
and Andrzej Trautman. I like to thank David Rowe and his colleagues at
Mainz University for their invitation to attend their Symposium and for
their splendid hospitality. \newline
Eugene Surowitz provided the figures.

elschucking@msn.com
\vfill

\end{document}